# Probability of Spontaneous Emission in Processes of the Above Threshold Ionization of Atoms

K.V. Ivanyan*

M.V. Lomonosov Moscow State University, Moscow 119991, Russia

Expression for the probability of the spontaneous emission of high-order harmonics when the emitter phases are synchronized and are obtained in the region where the multiphoton approximation is applicable to the description of the ionization of an atom (where the adiabaticity parameter satisfies $\gamma > 1$). The dependence of this probability on the main parameters of the pump wave and the atomic medium is established. Criteria for observing emission are formulated with the consideration of phase locking.

## 1. INTRODUCTION

The generation of harmonics at frequencies corresponding to an odd number of photons of an ionizing laser wave has been investigated in several experirnental [1-4] and theoretical [5-11] studies. The basic experimental laws governing the dependence of the intensity $I_s$ (s is the harmonic order) on the atom density in the medium being ionized, the focusing of the laser wave, the volume of the region where the atom beam interacts with the wave, etc. have been established.

One of the special features of the generation of high order harmonics is the nonlinear dependence of $I_s$ on the atomic density $n_a$ of the medium established in [3]. This result was naturally related to the phenomenon of phase synchronism during harmonic generation. In [9] values of the radiated intensities of different harmonics were obtained as a function of the intensity of the ionizing wave (Nd:YAG laser, $\lambda = 1064$ nm, intensity range from $5 \times 10^{12}$ to $5 \times 10^{13} W/cm^2$) by numerically solving Maxwell's equation in a nonlinear medium of Xe atoms. A comparison of the results of the calculations in that paper with the experimental data in [3] revealed generally good agreement.

In [11] and [12] the generation of high-order harmonics was studied using the analytical approach previously developed in [13] to describe the effects of the above threshold ionization of atoms. This method is based on a multiphoton mechanism for the ionization of an atom, for which the Keldysh adiabaticity parameter satisfies $\gamma > 1$. It is assumed that after the birth of a photoelectron at the threshold, it gathers additional energy as a result of repeated rescattering on the Coulomb potential of the residual parent ion, which is accompanied by the absorption of quanta of the field. In particular, the main laws governing the harmonic spectrum (the plateau and the cutoff regions) were described within this approach in [11], and the dependence of the order so of the cutoff harmonic on the intensity I of the laser wave was established. Numerical evaluations of so from the equations in [11] showed that good agreement both with the experimental results in [1] and [4] and with the theoretical calculations of other investigators [14,15] is observed in the region where the theory is applicable. The other aspects are presented in [17-70].

In the present work expression is obtained for the probability of the spontaneous emission of high-order harmonics under phase-locking conditions. The dependence of this probability on the basic parameters of the pump wave and the atomic medium is established.

-----------------

*k.ivanyan@yandex.com

## 2. BASIC EQUATIONS

We start out from the assumption that harmonic generation is directly related to the above-threshold ionization of atoms. Under this approach the generation mechanism is as follows. The low harmonics are emitted on transitions from excited bound states to the ground state of an atom or as a result of free-free transitions between photoelectron states in the continuum. The high-order harmonics arise in processes involving the direct spontaneous recombination of photoelectrons from highly excited states to the ground state of an atom.

We use the expression (h = c = 1 )

$$\widehat{V}(t) = \frac{e}{m_e} \mathbf{A}(t) \cdot \mathbf{p} + \frac{[eA(t)]^2}{2m_e} \tag{1}$$

to define the interaction operator of an electron with the pump wave field. The vector potential of the wave is given by the classical expression

$$\mathbf{A}(t) = \frac{A_0}{2} \mathbf{e} \left[ e^{i(\omega t - \mathbf{k} \cdot \mathbf{r})} + c.c. \right], \tag{2}$$

where $A_0$ is the amplitude of the vector potential, $\mathbf{k}$ and $\omega$ are the wave vector and the frequency of the wave, and e is the polarization unit vector (we assume that the wave is polarized linearly along the z axis: $\mathbf{e} = \mathbf{e}_z$ ).

The amplitude of the probability of recombination of the system to the ground state of the atoms with the emission of photons having the wave vector $\mathbf{K}$ and the frequency $\Omega$ at the time *t* is given by the expression [12]

$$\mathbf{A}_\Omega(t) = \tilde{A}_0 e A_{0\Omega} \sum_j \exp[i(s\mathbf{k} - \mathbf{K}) \cdot \mathbf{R}_j] \xi^*(s\omega - \Omega) \exp[i(s\omega - \Omega - i\tilde{\alpha})t]. \tag{3}$$

The following notations were used in (3):

$$\tilde{A}_0 \frac{12}{\sqrt{2\ln 4}} \alpha(\mathbf{e}_\Omega \mathbf{e}) \sqrt{\frac{\varepsilon \omega^4}{Ry^5}} J_1(z') J_{n_0/2-1}(z) \left( \sum_{n=1} \Phi(n) \sum_{q,q'=1} (-1)^q J_{2q}(z') J_{2q'}(z) \right),$$

where

$$\Phi(n) = \frac{\exp[-n/\ln(4n)]}{\sqrt{n}[1+(n\omega/Ry)^2]} \left[ \frac{Ry}{\omega} \left( \frac{ez'\ln(4n)}{4n} \right)^2 \right]^{n/2}$$

is a function which describes the envelope of the maxima in the spectra of the photoelectrons appearing as a result of above-threshold ionization; $s = n_0 + n + 2q + 4q'$ gives the harmonic order in the emission spectrum; $n_0 = \langle \tilde{I}_0/\omega + 1 \rangle$ [here $\tilde{I}_0 = I_0 + U_p$ is the binding energy of an electron in an atom with consideration of the mean vibrational energy in the field of the pump wave $U_p = (eE_0)^2/4m_e\omega^2$ where $E_0$ is the amplitude of the electric field strength in the wave and <...> is the whole part of the number]; $A_{0\Omega}$ and $\mathbf{e}_\Omega$ are the amplitude of the vector potential and the polarization unit vector of the emitted wave with a frequency $\Omega$; $\varepsilon = n_0\omega - \tilde{I}_0$ is the amount by which the energy of no photons of the wave exceed the ionization threshold (to be specific, we assume that $n_0$ is even);

$$z' = 2eE_0\tilde{\lambda}/\sqrt{2m_e\omega}, \quad z = (eE_0\tilde{\lambda})^2/8m_e\omega = U_p/2\omega$$

are dimensionless parameters, which specify the intensity of the interaction of an electron with the ionizing wave in accordance with the operator (1); $\tilde{\lambda} = 1/\omega$; $J_m(x)$ is a Bessel function; $\mathbf{R}_j$ is the radius vector of the j-th atom (residual ion) and the summation over j in (3) is carried out over all the atoms in the volume where the medium interacts with the pump wave; $\alpha = e^2/\hbar c$ is the fine-structure constant; $\xi^* = P/x + i\pi\delta(x)$; $Ry = m_e e^4/2\hbar^2 = 13.6$ eV ; the parameter $\tilde{\alpha} \approx 0$ corresponds to the adiabatic switching on of the wave field as $t \to -\infty$.

Expression (3) leads to a formula for the probability of a transition to a partial final state per unit time

$$\frac{d}{dt}|A_\Omega(t)| = \tilde{A}_0^2(eA_{0\Omega})^2 \left|\sum_j \exp[i(s\mathbf{k} - \mathbf{K})\cdot\mathbf{R}_j]\right|^2 2\pi\delta(s\omega - \Omega). \tag{4}$$

Equation (4) is the basis for further calculations and is used to derive the probabilities of both the spontaneous and induced emission of harmonics.

### 3. PROBABILITY OF SPONTANEOUS EMISSION

In the case of the spontaneous emission of the s-th harmonic of the frequency $\omega$ ($\Omega = s\omega$), the quantity $eA_{0\Omega}$ in (4) should be replaced by the expression $eA_{0\Omega} = \sqrt{8\pi\alpha/\Omega V}$, where V is the normalized volume of the spontaneous emission field. When the transition probability is obtained, the delta function in (4), which gives the energy conservation law, is replaced by integration over the statistical weight of the photon emitted with the parameters $\mathbf{K}$ and $\Omega$. As a result the probability of the spontaneous recombination of the system to the ground state of the atoms per unit time is given by the expression

$$w_{sp}^{(s)} = 4\alpha\tilde{A}_0^2 s\omega \int_{(4\pi)} \left|\sum_j \exp[i(s\mathbf{k} - \mathbf{K})\cdot\mathbf{R}_j]\right|^2 d\Omega_\mathbf{K}. \tag{5}$$

in which the integration is carried out over the scattering directions of the photon with the parameters $\mathbf{K}$ and $\Omega$.

The ensuing calculations are easily performed in the continuous-medium approximation (the criterion is formulated below), in which summation over the atoms within the interaction volume leads to the following result

$$\left|\sum_j \exp[i(s\mathbf{k} - \mathbf{K})\cdot\mathbf{R}_j]\right|^2 \approx \left(\frac{\pi}{6}\right)^2 (V_{int} n_a)^2 \left[\frac{2J_1(v)}{v}\right]^2 \frac{\sin^2 u}{u^2}$$

The following notations were introduced in (6): $V_{int} = \pi\rho_0^2 l$ is the volume of the region where the atomic medium interacts with the laser radiation [here $\rho_0$ is the radius of the focus in the plane with the coordinate x=0, l is the longitudinal dimension of the interaction region in the direction of propagation of the wave, which is specified by the condition l = min (L,d), where L is the confocal parameter and d is the diameter of the atomic beam aimed transversely toward the wave],

$$v = \frac{2\pi s}{\lambda_0} \rho_0 \theta, \quad u = \frac{\pi s}{2\lambda_0}(\theta^2 - \theta_0^2)l, \quad (6)$$

where $\lambda_0$ is the laser wavelength in a vacuum and $\theta_0^2 \equiv 2|\Delta n|$ ($\Delta n = n_\omega - n_\Omega$ is the difference between the refractive indices of the medium for waves of the respective frequencies; as the results of the numerical calculations in [9] under the conditions of the experiment in Ref. 3 showed, the dominant contribution to $\Delta n$ is made by the free electrons).

The expression (6) is valid provided $|s\mathbf{k} - \mathbf{K}|a \ll 1$, or, with consideration of the parameter $\Delta n$ just introduced,

$$\frac{|\Delta n|a}{\lambda_0/s} \ll 1$$

(a is the mean distance between atoms of the medium; in the continuous-medium approximation). The value of $|\Delta n|$ can be determined using the expression $|\Delta n| = \omega_p^2/\omega^2$ where $\omega_p^2 = 4\pi n_i e^2/m_e$ is the plasma frequency of the ionized medium ($n_i$ is the ion density in the medium, which is equal to the electron density: $n_i = n_e$).

As follows from (5) and (6), the angular density of the intensity of the spontaneous emission is given by the product of two diffraction factors:

$$\frac{dI_s}{d\omega d\Omega_{\mathbf{K}}} \propto \left[\frac{2J_1\left(\frac{2\pi s}{\lambda_0}\rho_0\theta\right)}{\frac{2\pi s}{\lambda_0}\rho_0\theta}\right] \frac{\sin^2\left[\frac{\pi s}{2\lambda_0}(\theta^2 - \theta_0^2)l\right]}{\left[\frac{\pi s}{2\lambda_0}(\theta^2 - \theta_0^2)l\right]^2}. \quad (7)$$

Far from saturation, where the concentration of photoelectrons is small, $\theta_0 \approx 0$ and the maxima of these functions coincide in the direction of propagation of the pump wave $\theta_0 = 0$. In the case of an ionizing wave of high intensity, in which the ionization of the medium is close to saturation ($n_i \leq n_e$), we have $\theta_0 \neq 0$, and the values of $\theta$ which give the maxima of the diffraction factors are different. For this reason, the total intensity of the spontaneous emission $I_s$ is sensitive to the value of $\theta_0$ and the ratio between this parameter and the angular widths of the diffraction factors.

In the case of $\theta_0 \approx 0$ (the smallness criterion is formulated below) the angular widths associated with the finite dimensions of the focal region in the longitudinal and transverse directions are defined, respectively, by the formulas

$$\Delta\theta_\parallel \approx \sqrt{\frac{2\lambda_0/s}{l}} \quad \text{and} \quad \Delta\theta_\perp \approx \frac{\lambda_0/s}{\rho_0}. \quad (8)$$

A comparison of the widths (8), in particular, for the parameters of the laser and the atom beam used in [3] leads to the inequality $\Delta\theta_\parallel > \Delta\theta_\perp$. For this reason, the total number $N_\gamma$ of photons of a given harmonic emitted during a pulse of the pump wave is given by the following relation, which follows from (5):

$$N_\gamma \propto \frac{\sin^2(\pi l / L_{coh})}{(\pi l / L_{coh})^2} \tag{9}$$

where

$$L_{coh} = \frac{2\lambda_0}{s\theta_0^2} = \frac{\pi m_e}{se^2 \lambda_0 n_i} \tag{10}$$

is the coherence length of the emitters in the direction of propagation of the waves, which depends on the harmonic order and the extent of ionization of the medium.

According to (9), phase locking is fully realized, if the condition $L_{coh} > l$ holds. This relation, which is written in the form of an inequality

$$n_i < \frac{\pi m_e}{se^2 \lambda_0 d}, \tag{11}$$

imposes an upper bound on the intensity of the laser radiation. No direct measurements of the density $n_i$ of the ions formed in the medium were performed in [3], but a numerical evaluation for the parameters in that paper gives $n_i < 10^{18} / s \cdot cm^{-3}$.

When the condition $\theta_0 \leq \Delta\theta_\perp$, holds, photons of the s-th harmonic are emitted in the direction of propagation of the pump wave within the solid angle

$$\pi(\Delta\theta_\perp)^2 \approx \frac{1}{\pi}\left(\frac{\lambda_0}{s\rho_0}\right)^2.$$

Here the total probability of spontaneous recombination of the system per unit time with the emission of photons of the s-th harmonic appears as a result of the integration in (5) and is given by the expression

$$w_{sp}^{(s)} = 2\pi\alpha^3 \left(\frac{2\pi}{\ln 4}\right)^2 \varepsilon \left(\frac{\omega}{Ry}\right)^5 \left[J_1(z')J_{n_0/2-1}(z)\right]^2$$

$$\times \frac{1}{s}\left(\sum_{n=1}\Phi(n)\sum_{q,q'}(-1)^q J_{2q}\left(z'\sqrt{n}\right)J_{2q'}(z)\right)^2 (\rho_0 l \lambda_0 n_a)^2. \tag{12}$$

If the condition for phase locking does not hold $L_{coh} \ll l$, the probability of spontaneous emission is proportional to the first power of $n_a$. The ratio of the probability (12) to the analogous probability obtained in the absence of phase locking of the emitters is given by the parameter

$$V_{int} n_a (\Delta\theta_\perp)^2 \approx \frac{l\lambda_0^2 n_a}{s^2}. \tag{13}$$

In particular, for the data from [3] this ratio has a sizable value ($\sim 10^7$).

Let us now consider the dependence of $w_{sp}^{(s)}$ (10) on the intensity I of the laser wave. This dependence is determined mainly by the factor $J_{n_0/2-1}(z)$, since the remaining factors scarcely vary with I. Bearing in mind that we have $z < 1$ and $n_0 \gg 1$ under the real conditions of the experiment in [3], from the asymptotic behavior of the Bessel function for a large index we find

$$w_{sp}^{(s)} \propto I^{n_0} \tag{14}$$

This dependence was established in [16] for the case of Xe atoms in the range of intensities of the laser wave where the photoionization process has a multiphoton character ($\gamma > 1$).

## 4. DISCUSSION OF THE RESULTS; CONCLUSIONS

The basic expressions for the probabilities of the emission of high-order harmonics were obtained under the assumption that the generation of these harmonics is directly related to the above-threshold ionization of atoms. A comparative evaluation of the amplitude of the oscillations $a_{osc}$ of a photoelectron in the external wave with a mean distance a between the atoms of the medium takes on fundamental significance in the context of this approach. Evaluating $a_{osc}$ from the formula $a_{osc} = eE_0/m_e\omega^2$ we can easily see that the value $a_{osc} \sim 10^{-7} cm$ for the laser wave intensities under consideration ($I \sim 10^{13} W/cm^2$) is an order of magnitude smaller than $a_{osc} \approx 10^{-6} cm$ ($n_a = 5 \cdot 10^{17} cm^{-3}$ in [3]). It follows from these estimates that the extraction of energy by a photoelectron after ionization occurs as result of rescattering on the Coulomb potential of its own residual ion.

For phase locking to occur during harmonic generation, a fixed phase must be maintained for the emitters over the course of the time for recombination of the system. Collisions of the atoms with one another are the main factor which can lead to phase relaxation. The mean free path of the atoms estimated via gas-kinetic arguments is $\lambda_{sc} \sim 10^{-4} cm$ for the parameters from Ref. 3. Accordingly, the mean time between two successive collisions is $\tau_{sc} \sim 10^{-8} s$ (for atoms with thermal velocities). This time is far greater than the pulse duration $\tau_p \approx 4 \times 10^{-11} s$ from [3], and, thus, recombination occurs in the photoelectron-ion system from states of electrons with a fixed phase appearing as a result of above-threshold ionization.

The probability (12) of the spontaneous emission of the s-th harmonic was obtained assuming a plane pump wave. This assumption is valid when the width of the angular spread of the wave vectors $\theta_f$, which is related to the focusing of the laser wave, satisfies the condition $\theta_f < \theta_0, \Delta\theta_\perp$. In the nearby diffraction zone (according to the conditions of the experiment in Ref. 3, the diameter d of the atom beam is less than the confocal parameter L) the maximum width of the angular spread can be evaluated using the formula

$$\theta_f \approx \frac{d\lambda_0^2}{(2\pi)^2 \rho_0^3}.$$

In a comparatively weak field ($\theta_0 < \theta_\perp$) the condition $\theta_f < \Delta\theta_\perp$ leads to an upper bound on the harmonic order s up to which the approximation used is valid:

$$s < \frac{(2\pi)^2 \rho_0^2}{d\lambda_0}.$$

For the parameters from [3] ($\rho_0 = 18\mu$, d= 1 mm, $\lambda_0 = 1064$ nm) this condition holds up to $s \approx 15$.

In a strong field ($\theta_0 > \Delta\theta_\perp$), in which the main cause of the breakdown of phase locking is the difference between the refractive indices $\Delta n$ of the pump wave and the emitted harmonic, a lower bound on the ion density follows from the condition $\theta_0 > \theta_f$:

$$n_i > \left(\frac{d\lambda_0}{2\rho_0^3}\right) \frac{m_e}{(2\pi)^3 e^2}. \tag{20}$$

This condition leads to the numerical estimate $n_i > 10^{16} cm^{-3}$ for the parameters from [3]. It is noteworthy that the estimates of the permissible values of $n_i$ following from (11) and (20) are compatible.

Breakdown of the phase locking of the emitters can be associated not only with the focusing effect, but also with nonmonochromaticity of the pump wave. In this case the longitudinal length $L_{eff}$ over which coherence of the emitters is maintained, is given by the relation

$$s \frac{\Delta\omega}{\omega} \frac{L_{eff}}{\lambda_0} \sim 1, \tag{21}$$

where $\frac{\Delta\omega}{\omega}$ is the frequency spread of the pump wave, which is determined by the pulse duration. For the data from [3], $L_{eff}$ is of the order of the longitudinal length of the focus $L \approx 4m$.

The dependence of the intensity $I_s$ of the s-th harmonic on the position of the atom beam relative to the center of the focus of the laser wave can be established using relations obtained in the present work. As follows from (5) and (6), this intensity is given by the dependence

$$I_s \propto I^{n_0} \frac{\sin^2(\pi l / L_{coh})}{(\pi l / L_{coh})^2}. \tag{22}$$

In the case of a weak wave ($L_{coh} \gg l$), as has already been noted, $I_s \propto I^{n_0}$. Therefore, displacement of the beam from the center of the focus results in a monotonic decrease in the intensity of the harmonics. In the case of a high-intensity wave, in which $L_{coh} < l$ holds at the center of the focus, this dependence has a different character: $I_s \propto I^{n_0}(L_{coh}/l)^2 \propto I^{-n_0}$. Under these conditions an increase in the distance of the beam from the center of the focus can be accompanied by an increase in $I_s$. This increase will occur up to the distances from the center of the focus at which the local intensity of the wave field reaches values corresponding to the condition $L_{coh} \sim l$. Further displacement of the beam toward the periphery will reproduce the picture obtained using perturbation theory: a monotonic decrease in $I_s$ with increasing distance from the center of the focus. These qualitative arguments were confirmed by the numerical calculations in [9].